\documentclass{jpp}
\usepackage{graphicx}
\usepackage{epstopdf, epsfig}
\usepackage{hyperref}

\shorttitle{Magnetothermodynamics}
\shortauthor{Kaur et. al.}

\title{Magnetothermodynamics: Measurements of the thermodynamic properties in a relaxed magnetohydrodynamic plasma}

\author{M. Kaur\aff{1}
	\corresp{\email{mkaur2@swarthmore.edu}},
	L. J. Barbano\aff{1}, E. M. Suen-Lewis\aff{1}, J. E. Shrock\aff{1}, A. D. Light\aff{1}, D. A. Schaffner\aff{2}, M. B. Brown\aff{1}, S. Woodruff\aff{3} and T. Meyer\aff{3}}

\affiliation{\aff{1}Swarthmore College, Swarthmore, Pennsylvania 19081 USA
	\aff{2}Bryn Mawr College, Bryn Mawr, Pennsylvania 19010 USA
	\aff{3}Woodruff Scientific Inc., Seattle, Washington 98103 USA}

\begin{document}

\maketitle

\begin{abstract}
We have explored the thermodynamics of compressed magnetized plasmas in laboratory experiments and we call these studies ``magnetothermodynamics". The experiments are carried out in the Swarthmore Spheromak eXperiment device. In this device, a magnetized plasma source is located at one end and at the other end, a closed conducting can is installed. We generate parcels of magnetized plasma and observe their compression against the end wall of the conducting cylinder. The plasma parameters such as plasma density, temperature, and magnetic field are measured during compression using HeNe laser interferometry, ion Doppler spectroscopy and a linear $\dot{B}$ probe array, respectively. To identify the instances of ion heating during compression, a PV diagram is constructed using measured density, temperature, and a proxy for the volume of the magnetized plasma. Different equations of state are analyzed to evaluate the adiabatic nature of the compressed plasma. A 3D resistive magnetohydrodynamic code (NIMROD) is employed to simulate the twisted Taylor states and show stagnation against the end wall of the closed conducting can. The simulation results are consistent to what we observe in our experiments. 
\end{abstract}

\section{Introduction:}
The equation of state (EOS) of an ideal gas as derived from statistical mechanics relates pressure, volume, and temperature of the gas. Charged particles in a plasma, particularly if there is a strong magnetic field, need not obey an ideal gas EOS. In such a case, one needs to consider the motion of charged particles both along and across the local magnetic field.
Measuring the EOS of a magnetized plasma is important for advancing fusion experiments as well as for understanding natural systems such as stellar winds. 

The true equation of state in the solar wind, for example, is still unknown. Solar wind plasmas are collisionless and in the supersonic expansion of such plasmas, an anisotropy between perpendicular temperature ($T_\bot$) and parallel temperature ($T_\parallel$) with respect to the direction of the background magnetic field is observed \citep{Kasper2003}. If the expansion of the solar wind plasma were to be regulated by an adiabatic equation of state, the expected drop in the proton temperature (particularly $T_\bot$) would be much faster with radius from the Sun \citep{Brown2015} than is measured at $1~AU$. This is because the volume of a shell of solar wind plasma increases as the square of the radius, i.e., $r^2$ and $1~AU$ is about $200$ solar radii. Instead, the effects of collisional age (time measured in units of mean time between collisions) \citep{Bale09} and kinetic instabilities (such as firehose and mirror) \citep{Kasper02} appear to regulate the temperature anisotropy but do not yet account for the thermodynamics of the expansion. 

Over the past decades, numerous experiments \citep{park14,Hurricane14,Hammer88, Hammer91, Molvik91} motivated by fusion applications have been performed to achieve a highly compressed plasma but little has been done to understand the thermodynamics of such systems. Some progress has been made in measuring the EOS of unmagnetized plasma in the context of inertial confinement fusion experiments \citep{park14,Hurricane14}. At the National Ignition Facility (NIF), a radial compression of an unmagnetized DT pellet resulted in a density increase by a factor of $1000$. The full NIF EOS is unpublished but includes radiation as well as compressive effects. Similar thorough studies in magnetized plasma have yet to be performed.

Early compression experiments on magnetized plasma were performed at the RACE facility in the 1990s \citep{Hammer88, Hammer91, Molvik91}. In these experiments, spheromak-type axi-symmetric plasmas were magnetically accelerated to $2500~ km/s$ along $6~m$ long coaxial electrodes and stagnated in a conical taper. Radial compression of a factor of three and magnetic field amplification by a factor of four were observed. In compression experiments on the Inductive Plasma Accelerator (IPA) device, two axi-symmetric field reversed configurations (FRCs) were merged and magnetically compressed to kilovolt ion temperature \citep{slough2011, Votroubek08}. Velocities of $250~ km/s$ were achieved in IPA. Finally, in liner implosion experiments at the Shiva Star facility, dense, initially-unmagnetized hydrogen plasma was compressed to megabar pressures inside a magnetically imploded aluminum spheres. A radial compression of a factor of three was observed in the Shiva Star experiments \citep{Degnan99}. However, an equation of state (EOS) was not reported in any of these experiments. As compression experiments in magneto-inertial fusion (MIF) \citep{Lindemuth83, Lindemuth2015,Lindemuth2017} come online, it will be important to understand the EOS of such magnetized plasmas.

Recently, compression experiments were carried out at the magnetized liner inertial fusion (MagLIF) experiment at Sandia \citep{Awe2013, Gomez2014, Slutz2012}. In these experiments, a $4.6~mm$ diameter cylinder of deuterium is premagnetized to $10~T$, then preheated by a laser pulse to $100~eV$ and imploded by the Z machine. Physical scale and temporal duration ($100~ns$) of these experiments are much larger than that of typical ICF experiments which significantly relaxes constraints on the driver. Final plasma parameters upon compression at MagLIF were as follows: plasma temperature increases to $4~keV$, magnetic field to $1000~T$, and DD neutron yield of around $2 \times 10^{12}$ was achieved. These are promising results that could be aided by well-established EOS.

In this paper, the experimental studies in which we have explored the thermodynamics of compressed magnetized plasmas (``magnetothermodynamics") using measured plasma parameters, described in \cite{ManjitPRE}, are presented in detail. 
In these experiments, we generate a parcel of magnetized, fully relaxed, non-axisymmetric plasma \citep{Taylor1986, Taylor74, Woltjer58, cothran_prl, gray_prl} and observe its compression against a conducting cylinder closed at one end. The plasma parameters are measured during compression and a PV diagram is constructed to identify instances of associated ion heating; we call these ``compressive heating events''. The local magnetohydrodynamic (MHD) and the double adiabatic (CGL) equations of state are tested during compression events, under several experimental conditions. Our aim for these studies is to know whether 
there is an equation of state that describes the ion compressional heating of such relaxed plasmas.
We find that the MHD EOS is not consistent with our observations, while the parallel CGL equation of state is valid for our data.
To understand the axial variation of the plasma parameters, a 3D resistive magnetohydrodynamic code (NIMROD) is employed, which shows a helical structure consistent with previous observations \citep{gray_prl}. A Taylor state was observed to stagnate against the end wall of the closed conducting can in simulations.

The remainder of the paper is organized as follows. In section II, we review the theory of equations of state for magnetized plasmas, both the MHD EOS and the so-called double adiabatic EOS. In section III, we describe the experimental set-up of the SSX plasma accelerator. In section IV, we discuss our three principal diagnostics: an ion Doppler spectrometer for sub-microsecond temperature measurements, a HeNe laser interferometer for density measurements, and a magnetic probe array for magnetic structure and time-of-flight measurements. In section V, we present our experimental results, and in section VI, our simulation results are described. A discussion and conclusion is presented in section VII.  
\section{Theory:}
Equations of state for magnetized plasmas fall into two categories.  In the collisional regime (i.e., $ \omega_{ci}\tau < 1 $, where $ \omega_{ci} $ and $\tau$ are the ion cyclotron frequency and Coulomb collision time, respectively), ions suffer multiple collisions before executing a Larmor orbit. In this case, magnetic field plays no role; MHD plasma exhibits an isotropic velocity distribution and can be treated as an ideal gas. The adiabatic EOS for such a plasma can be written
\begin{equation}
\frac{\partial}{\partial t}\left(\frac{P}{n^{\gamma}}\right) = 0, \label{mhd_eos}
\end{equation}
where $P$ and $n$ are the plasma pressure ($P = nk_B T$) and density, respectively; $T$ is the plasma temperature. 
$\gamma$ is the ratio of heat capacity at constant pressure to heat capacity at constant volume. Its value is given by $ \gamma = (f + 2)/f $, where $f$ is the number of microscopic degrees of freedom. Thus, for an MHD plasma with three degrees of freedom, $ \gamma = 5/3 $.

If the collision rate is low (i.e., $ \omega_{ci}\tau > 1 $), ions execute many Larmor orbits before suffering a collision. In this case, the perpendicular ($P_\bot = nk_B T_\bot$) and parallel pressure ($P_\parallel= n k_B T_\parallel$) with respect to the direction of the background magnetic field need not be the same. Because of this anisotropy, the MHD equation of state no longer remains valid. To account for such a situation, Chew, Goldberger and Low proposed a modified set of adiabatic EOS, commonly known as double adiabatic or CGL equations of state \citep{CGL}:
\begin{equation}
	\frac{\partial}{\partial t}\left(\frac{P_\bot}{nB}\right) = 0 \label{<cgl_perp>},
\end{equation}
\begin{equation}
	\frac{\partial}{\partial t}\left(\frac{P_\parallel B^2}{n^3}\right) = 0 \label{<cgl_par>}.
\end{equation}
Eq. (\ref{<cgl_perp>}) is related to the constancy of the first adiabatic invariant $ \mu = W_\bot/B$ and Eq. (\ref{<cgl_par>}) is related to the constancy of the second adiabatic invariant, $ \mathcal{J} = v_\parallel L $; where  $W_\bot$, $v_\parallel$ and $L$ are the perpendicular kinetic energy with respect to magnetic field, parallel velocity with respect to magnetic field, and the length of the guiding center along the field lines, respectively. 
Equation (\ref{<cgl_par>}) follows from the assumption of constant magnetic flux along a flux tube and fixed number of particles.
While deriving the Equation (\ref{<cgl_perp>}) and (\ref{<cgl_par>}), the plasma oscillations have been neglected as for these plasmas, the plasma frequency is very large compared to the ion cyclotron frequency. 

\section{Experimental set-up:}
\begin{figure*}
	\includegraphics[width=\textwidth]{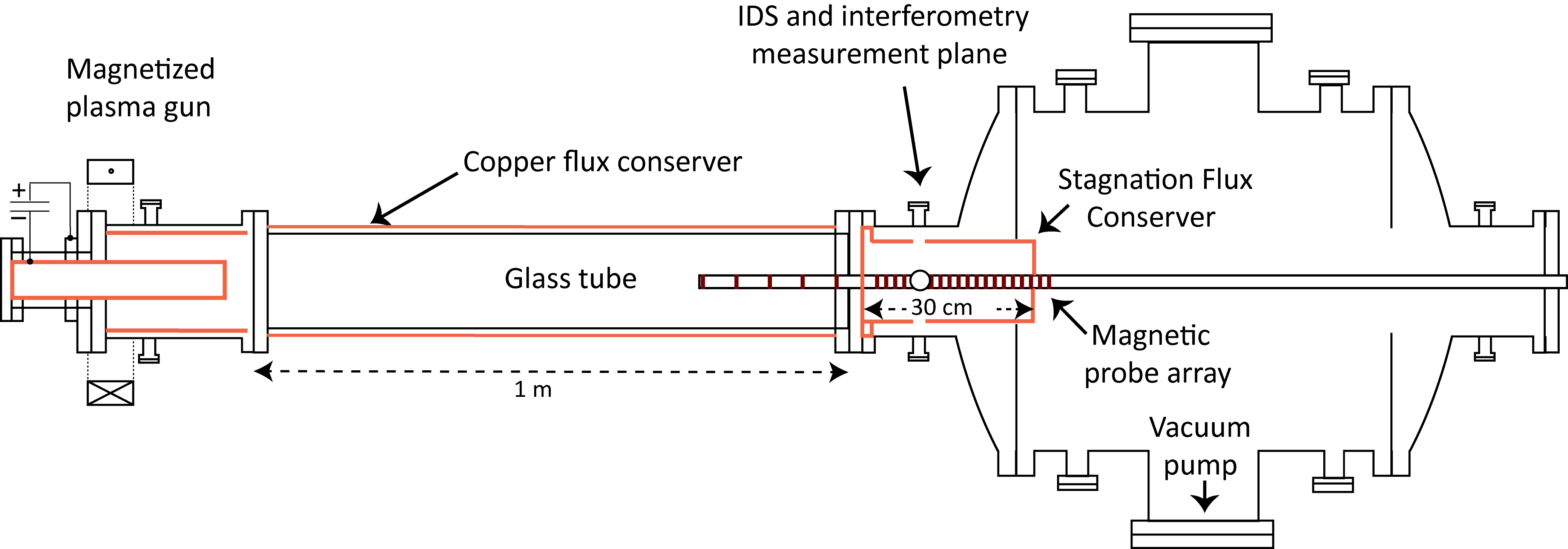}
	\caption{\label{fig:schematic} A schematic of the experimental set-up. A glass tube has been added in between the gun and the stagnation flux conserver (SFC) and is covered with a copper flux conserving shell. A Permendur magnet rod is installed at the center of the inner electrode shell. All the three principal plasma diagnostics are located in the SFC at the location of the circle. $T_i$ is measured using ion Doppler spectroscopy along the vertical chord and $n$ is measured using HeNe laser interferometry along a horizontal chord. The long $\dot{B}$ probe array is aligned along the axis of the SFC. The red lines on the magnetic probe array represent the locations of the $\dot{B}$ probes.}
\end{figure*}

In these experiments, we produce parcels of magnetized plasma using a coaxial magnetized plasma gun, which is located at one end of the linear Swarthmore Spheromak eXperiment (SSX) device, as shown in Fig.~\ref{fig:schematic}.
The diameter of the inner electrode of the gun is $6.2~cm$ and the outer electrode (i.e., flux conserver) diameter is $15~cm$. More details about the plasma gun can be found in earlier publications \citep{Brown2015, Geddes1998,Brown2014PSST}. At the other end of the linear chamber, a closed, tungsten-lined copper can, referred to here as a stagnation flux conserver (SFC), is installed. The SFC is $\sim 30~cm$ long and has the same inner diameter as that of the outer electrode. A $\sim1~m$ long quartz tube (diameter = $15~cm$) is installed in between the gun and the SFC to accommodate theta-pinch coils for our future experiments. For the present studies, the glass tube has been covered with a copper flux conserving shell having long magnetic soak time ($> 200~ \mu s$) to provide flux conservation to the magnetized plasma. We maintain a good vacuum using a cryopump. After each vacuum break, the gun and the device are cleaned using He glow discharge to maintain a low-impurity plasma.

Tungsten-lined coaxial electrodes are cleaned by producing a He glow discharge before experiments to obtain good wall conditions. A strong magnetic field ($\backsim 1~T$) is generated in the inner electrode using an external electromagnet and a strong ferromagnetic core. Hydrogen gas is puffed into the annular region between the two electrodes using gas puff valves. A voltage pulse ($\approx 4~kV,~8~kJ$) is applied between the two electrodes which ionizes the gas and causes a high current ($\backsim 100~kA$) to flow through the plasma. $\mathbf{J} \times \mathbf{B}$ forces accelerate the plasma out of the gun and a toroidal self-consistent magnetic object, called a spheromak \citep{Geddes1998}, is formed. 

The neutral hydrogen gas density (and hence, plasma density) can be controlled by varying the delay between the gas puff valve and gun. The gas puff valves are similar to those described by Thomas \textit{et. al.} \citep{Thomas1993}. Using this scheme, a proper delay will provide only as much gas as is necessary to form the parcels of magnetized plasma. If the delay is too short, i.e., the gun fires too early (in our case, $300 ~\mu s$ after the gas puff), there will not be enough hydrogen gas to ionize and this may result in sputtering of the electrode surface. If the delay is too long (e.g., the gun fires $750 ~\mu s$ after the gas puff), due to the thermal speed of the gas atoms, some hydrogen gas would leave the gun and enter the glass tube. In such a case, the spheromak will form by ionizing the hydrogen gas present in the annular region; but as soon as it enters the glass tube, due to its interaction with the neutral hydrogen gas, the spheromak will cool. For most of our experiments, the delay between the gas puff valve and the gun has been set between $500~\mu s$ to $600~\mu s$.

The toroidal structure continues to move away from the gun after its formation; the leading edge of the Taylor state moves at $60~km/s$ whereas the bulk plasma moves at $v\geq 40~km/s$. The copper sheet wrapped around the glass tube acts as a flux conserver with long aspect ratio for which the predicted minimum energy state is close to the infinite cylinder solution. Therefore, the toroidal structure tilts and relaxes to a non-axisymmetric twisted, force-free, minimum energy state, called a Taylor state, within few Alfv\'en times. This minimum energy state has been well-characterized in previous studies \citep{Taylor1986, Taylor74, Woltjer58, cothran_prl, gray_prl, Schaffner2014,Brown2014PSST}. The inertia of the Taylor state carries it to the other end of the device into the SFC, where the object stagnates and compresses against the end wall; this is where we perform measurements on the compressed plasma.
\section{Diagnostics:}
For these experiments, we rely on three principal diagnostics. While selecting these diagnostics, our goal is to be minimally invasive. For measuring the equations of state, it is important that the measurements perturb the plasma as little as possible. For measurement of ion temperature and plasma density, we use non-invasive diagnostics, ion Doppler spectroscopy and HeNe laser interferometry, whereas for the measurement the local background magnetic field, we use a long, axial $\dot{B}$ probe array. All these three diagnostics have a high frequency response, $\geq 1~MHz$, which is sufficient to resolve the physics of processes lying in the proton cyclotron frequency regime, $\sim 1.5~MHz$ for $ 1~kG$ magnetic field in our experiments. 
\subsection{Ion Doppler spectroscopy}
\begin{figure}
	\includegraphics[width=0.75\textwidth]{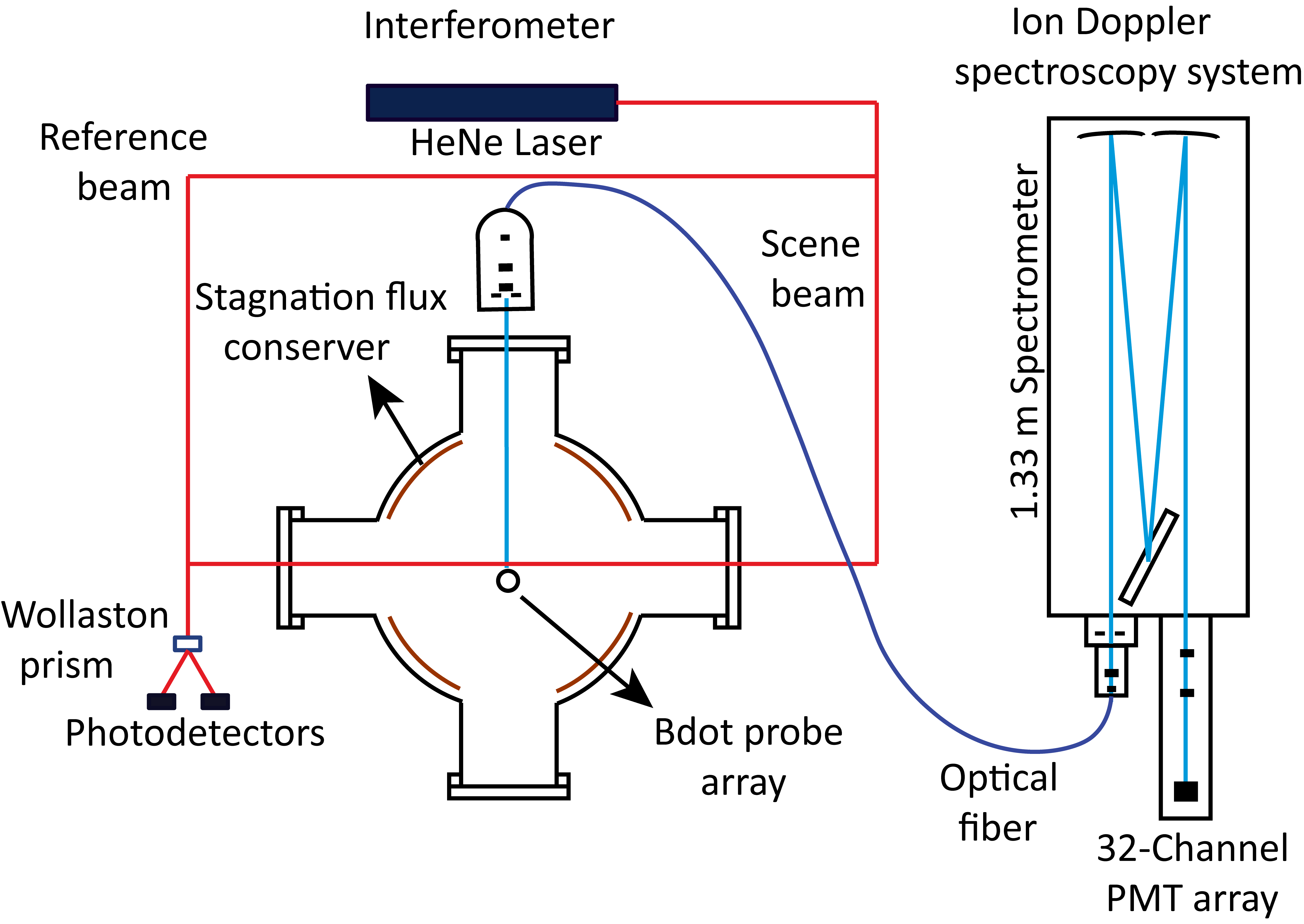}
	\caption{\label{fig:side_view} A cross-sectional view of the set-up along with diagnostics. In ion Doppler spectroscopy, $C_{III} $ line ($229.687 nm$) is dispersed using an echelle grating and recorded at 25th order. In laser interferometry, the interference output is passed through a Wollaston prism which generates two outputs $90^{\circ}$ out-of-phase with each other. The circle indicates the position of the $\dot{B}$ probe array, slightly off-centered.}
\end{figure}

\begin{figure}
	\includegraphics[width=0.75\textwidth]{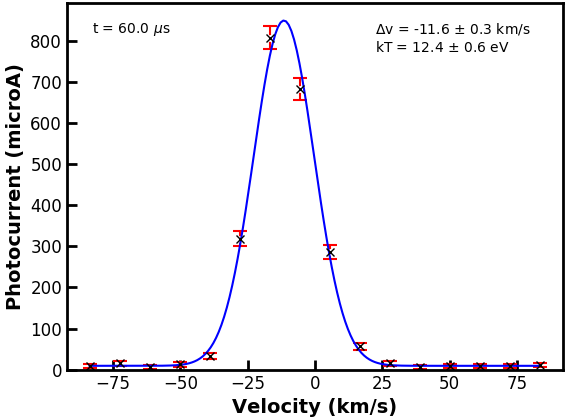}
	\caption{\label{fig:Line_shape_60mus} A typical line-shape obtained from the ion Doppler spectroscopy at $60~ \mu s$ using 16 channels on PMT array. The cross-marks in the black color correspond to the photocurrent and the red color bars represent their corresponding errorbars. The blue line represents the gaussian fit to the IDS data and the ion temperature obtained from the half-width-full-maxima of this gaussian lineshape is $\sim 12~eV$.}
\end{figure}

To measure ion temperature with sub-microsecond time resolution, we make use of ion Doppler spectroscopy (IDS). 
Our IDS system features a $ 1.33~m $ (Czerny-Turner) spectrometer which makes use of an echelle grating (with $ 316$ grooves/mm) to achieve high spectral resolution and a 32-channel photomultiplier tube (PMT) array for fast time response \citep{cothran_rsi}. The plasma light is collected along a vertical chord using a telescope, as shown in Fig.~\ref{fig:side_view},%
dispersed on the echelle grating, and is recorded using the PMT array. Our current studies are focused on the evolution of $229.687~nm$ emission line from $CIII$ impurity ions present in our plasma \citep{Chaplin2009} which we observe at $25^{th}$ order over a lifetime of $\simeq 100~\mu s$. The proton-impurity ion equilibration time is faster than $1~\mu s$ and the instrument temperature is $3-5~eV$. We are using 16 channels available on the PMT array. A typical lineshape at $60~ \mu s$ is shown in Fig.~\ref{fig:Line_shape_60mus}.%
The ion temperature obtained from this lineshape is $12~eV$ and the shift in the gaussian fit along the negative x-axis indicates a vertical velocity of $11.6~km/s$. The time resolution of these measurements is $\backsimeq 1~\mu s$ and is sufficient to observe any magnetohydrodynamic phenomena.

\subsection{HeNe laser interferometry}
\begin{figure}
	\includegraphics[width=0.75\textwidth]{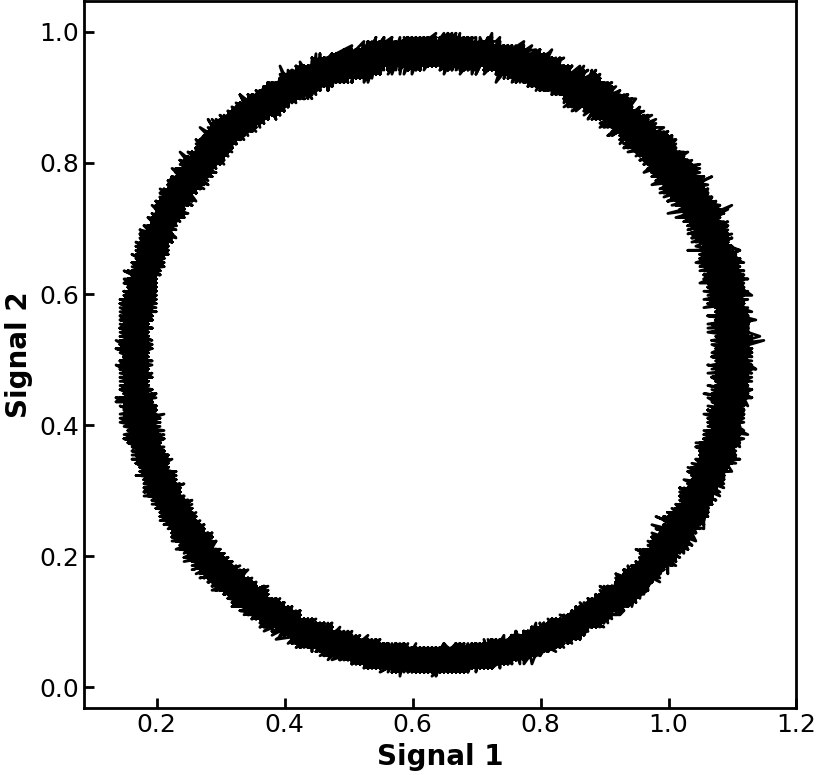}
	\caption{\label{fig:Density_Envelope} A typical envelope obtained from the interferometer detectors to be used in calibration for determining the electron density.}
\end{figure}

The line-averaged density of the Taylor state is measured along a horizontal chord using a 632.8 nm HeNe laser quadrature interferometer. The interferometer utilizes a modified Mach-Zehnder configuration and employs a Wollaston prism \citep{Brown2014PSST}. Phase shift ambiguities associated with normal interferometers are resolved by circularly polarizing the reference beam. Interference signal is passed through the Wollaston prism which generates two outputs $90^{\circ}$ out of phase with each other to enable density measurements during each shot. The intensity of each Wollaston output is measured by two photo-detectors. During the alignment of the interferometer, we make sure that the two output of the Wollaston prism are exactly $90^{\circ}$ out of phase. A typical envelope obtained from the interferometer is shown in Fig.~\ref{fig:Density_Envelope}.%
The initial phase of the measurements is removed from the two signals by considering the fact that at the beginning of the measurements, density of the plasma should be zero. 
As the timescale of the mechanical vibrations of the system ($ms$) is much larger than the timescale of the measurements ($\mu s$), the changes in phase due to the mechanical vibrations 
can be approximated as linear. This linear change in the data is removed by fitting a straight line to the data points in the range $t<25~\mu$s, as the Taylor states consistently arrive at the interferometer after this time.
The interferometry chord lies in the same plane as the IDS chord.

\subsection{$\dot{B}$ probe array}
Along with these two diagnostics, we use a long $\dot{B}$ probe array (encased in a domed quartz glass tube of outer diameter $1~cm$) along the axis of the SFC. The $\dot{B}$ probe array serves to measure the velocity and structure of the Taylor states and has been divided into two sections accordingly: the front and the rear section. Front section has three equally spaced ($14.2~cm$) single turn probes axially located inside the glass tube for measuring the Taylor state velocity using a time of flight (ToF) technique. 

The rear section of the long probe has densely spaced ($1.5~ cm$), single turn probes. These probes measure the magnetic field in two orthogonal directions, perpendicular to the flux conserver axis. These two-directional probes are located inside the SFC and are employed for determining the Taylor state structure. This section of the long probe also has few three-directional probes, which are used to measure the all three components of the local vector magnetic field and hence, to measure the local $\beta$ parameter. A three-directional $\dot{B}$ probe is nearly co-located with the intersection of IDS and interferometry chords. 

Each probe in the long probe array is carefully calibrated using a Helmholtz coil set driven by the same pulsed power supply that energizes the magnetized plasma gun. The $\dot{B}$ probe data is recorded at a high cadence ($65~MHz$) using 14 bit D-Tacq digitizers 
and is numerically integrated using the trapezoidal rule to obtain the absolute magnetic field values.
\section{Experimental Results:}
\begin{figure}
	\includegraphics[width=0.75\textwidth]{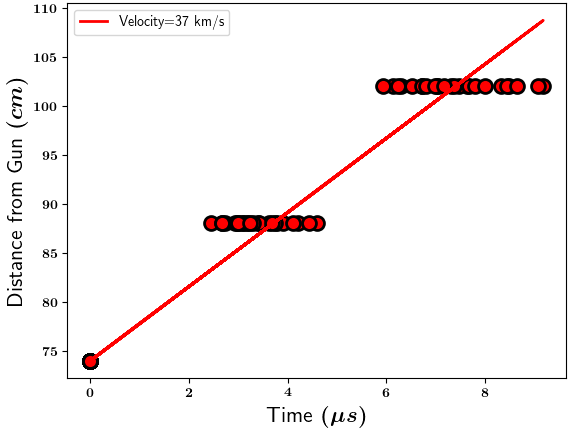}
	\caption{\label{fig:ToF_velocity} Velocity of the Taylor state using time of flight method from three $\dot{B}$ probes. The bulk Taylor state velocity is $\sim 37~km/s$ whereas its leading edge moves at $\sim 60~km/s$.}
\end{figure}

The bulk velocity of the Taylor state is measured using the front three $\dot{B}$ probes present in the glass tube on the axial long probe. 
The velocity of the leading edge of the Taylor state is determined by recording the time when the $\dot{B}$ probes start showing magnetic signal due to the plasma. In this way, the leading edge of the Taylor state is found to be moving at $\approx 60 ~km/s$.  
For determining the Taylor state bulk velocity, we track one particular structure in each probe data and record the time taken by the structure to reach the different probes. 
Using the time lag and the known distance between the probes, time of flight velocity of the bulk plasma is obtained. The bulk velocity of the Taylor state from time of flight is found to be $\sim 37 km/s$ and is plotted in Fig.~\ref{fig:ToF_velocity}.
The Taylor state flow speed is consistent with free expansion.
The Alfv\'en speed is $> 60~km/s$; therefore, the plasma (moving at $\sim 40~km/s$) stagnating against the end of the SFC is not expected to produce shocks.

\begin{figure}
	\includegraphics[width=0.75\textwidth]{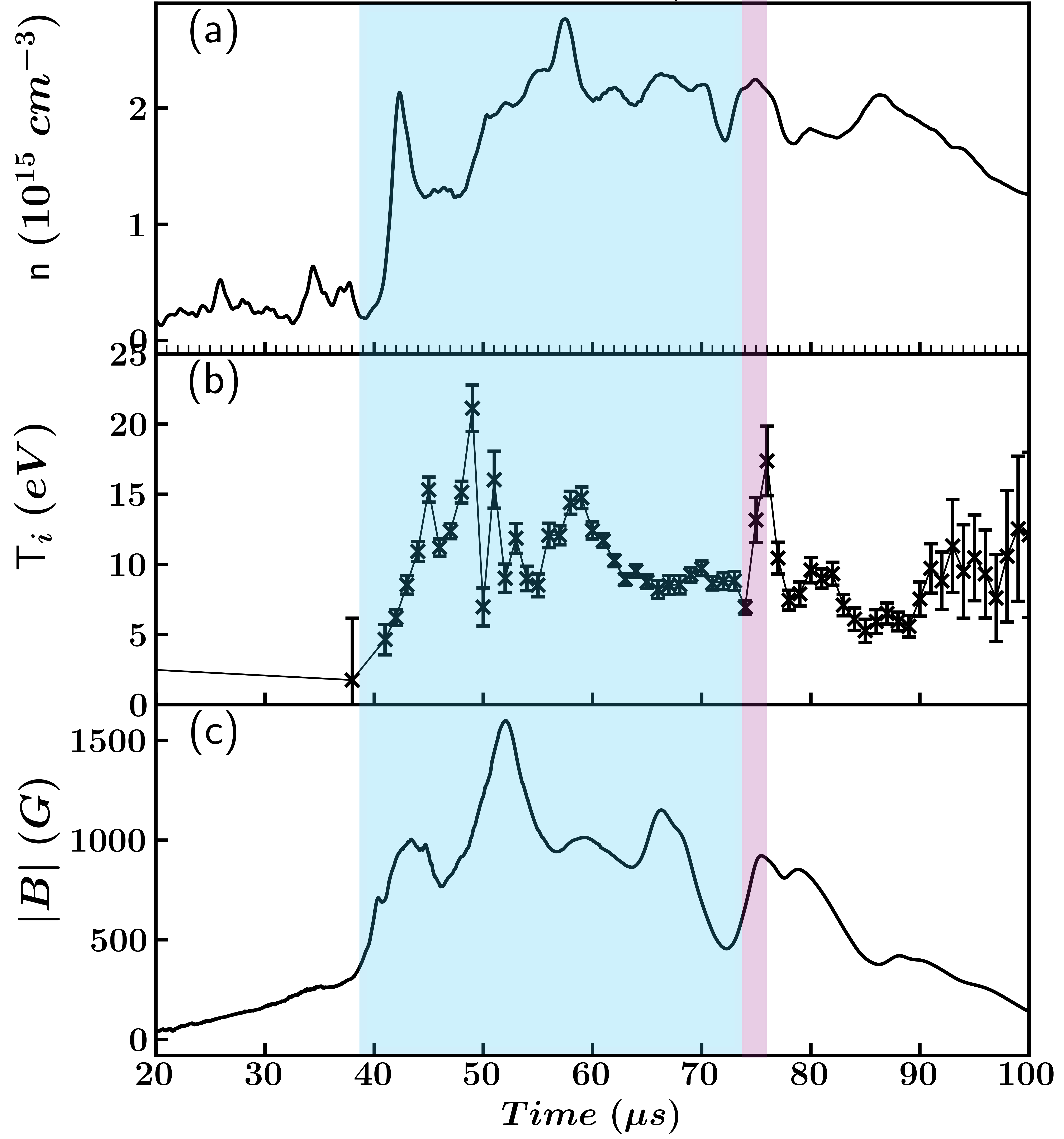}
	\caption{\label{fig:plasma_param} A typical time trace of (a) plasma density, (b) ion temperature, and (c) magnetic field measured inside the stagnation flux conserver. The error bars for $T_i$ are indicated by the vertical bars at each time value whereas the uncertainty in $n$ and $B$ is $< 10\%$. The 10\% uncertainty accounts for the errors due to a high frequency buzz present in the interferometer detector signals for density (also can be seen in Fig.~\ref{fig:Density_Envelope}) and due to the numerical integration in the magnetic field. Taylor state enters the SFC at $\approx 40~\mu s$ (indicated by the blue bar) and then compresses against the closed end of the SFC accompanied by a rise in $T_i$ at $40 \mu s$, $55~\mu s$ and $75~\mu s$ (indicated by the pink bar).}
\end{figure}

As mentioned in section III, all other measurements are performed on the compressed plasma inside the SFC. A typical set of time traces of plasma density, ion temperature and magnetic field magnitude measured inside the compression volume (SFC) is shown in Fig.~\ref{fig:plasma_param}%
. The leading edge of the Taylor state moves at $\approxeq 60 ~km/s$ and reaches the SFC at $40~\mu s$, indicated by blue color in Fig.~\ref{fig:plasma_param}. We observe a rise in plasma density and ion temperature for time windows spanning $40 - 48~\mu s$, $55 - 60~\mu s$ and $73 - 75~\mu s$ corresponding to compression against the back end of the SFC. 
The exact mechanism behind the increase in thermal pressure ($p =nT$) at these different times is not fully understood. However, we believe that the initial two events correspond to a compression of the Taylor state after it reaches the stagnation flux conserver. Whereas, the heating event occurring late in time might correspond to the compression of the Taylor state due to the arrival of an additional plasma entering the stagnation flux conserver. We occasionally see an additional plasma plume generated and enter the stagnation flux conserver late in time. 

The time trace of magnetic field magnitude in panel (c) of Fig.~\ref{fig:plasma_param} is obtained from the three-directional $ \dot{B} $ probes co-located with the IDS and interferometry chords.
Electron temperature, $T_e$ has been reported earlier in SSX Taylor states \citep{Brown2014PSST}. 
We have found that $T_e$ is around $\sim10~eV$ and is much less temporally dynamic than $T_i$. Since $T_e$ is essentially constant during a compression event, it does not contribute towards the EOS and has not been considered for further analysis. 
Electron thermal speed is $>10^8 ~cm/s$, but electron-electron collision mean-free-path is $\sim0.5~mm$, which indicates that the parallel electron transport is collisional and diffusive. Nonetheless, during a compressive heating event, electrons rapidly diffuse away from the measurement volume while the ions stay more localized.

\begin{figure}
	\includegraphics[width=\textwidth]{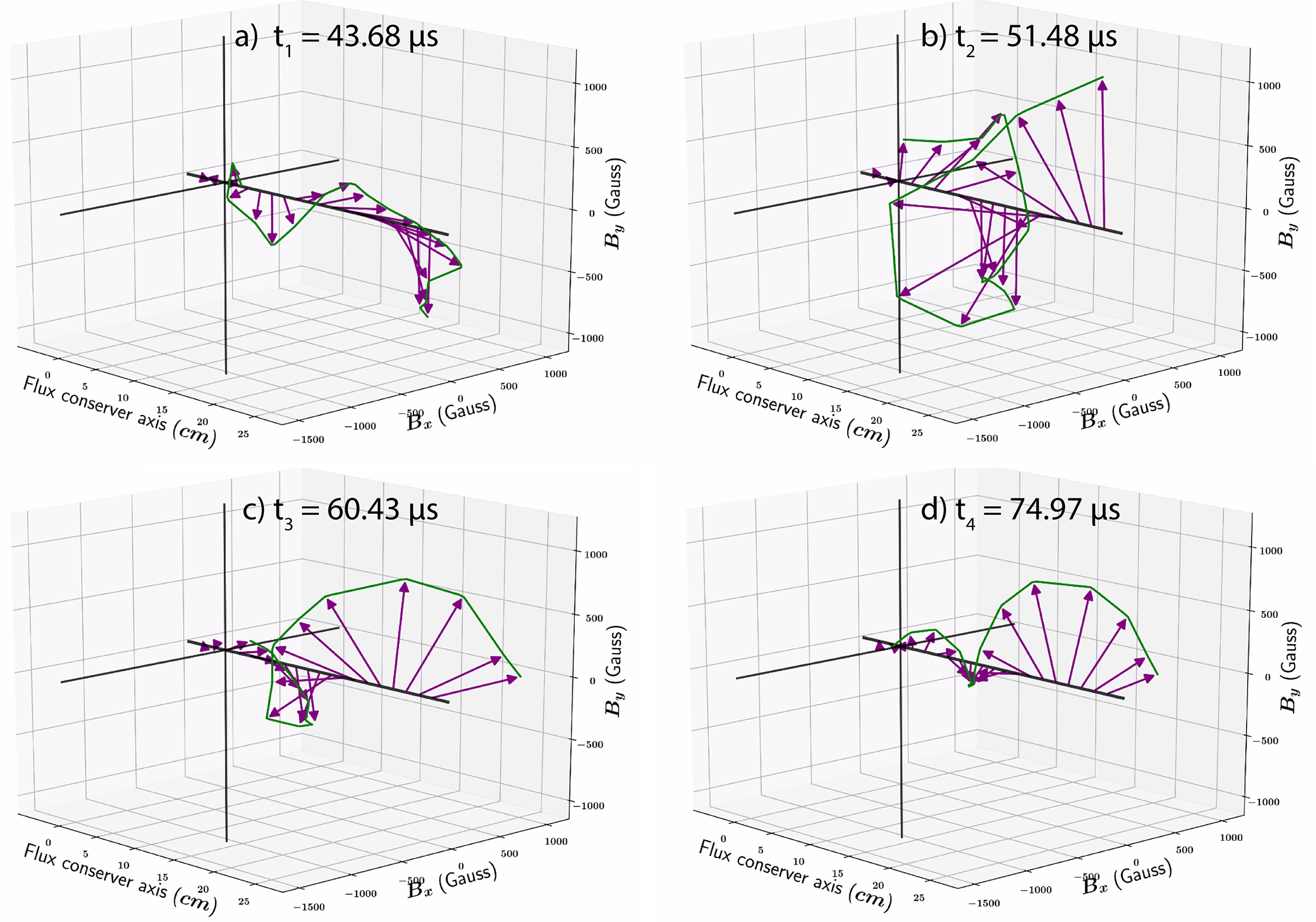}
	\caption{\label{fig:helix} A typical helical structure from the $\dot{B}$ probe array along the axis of the SFC at different time instances (for the same shot as Fig.~\ref{fig:plasma_param}): (a) $t_1 = 43.68~\mu s$, (b) $t_2 = 51.48~\mu s$, (c) $t_3 = 60.43~\mu s$, and d) $t_4 = 74.97~\mu s$. The two solid black lines represent the back end of the flux conserver. The twisted Taylor state enters the SFC from right hand side and gets compressed against the back end of the SFC.}
\end{figure}

The magnetic field vectors from the $\dot{B}$ probe array along the SFC axis are plotted at different times, as shown in Fig.~\ref{fig:helix}%
.   
The two solid black lines represent the back end of the flux conserver. The Taylor state enters the SFC from the right. At $43.68~\mu s$, one lobe of the helical structure is seen inside SFC (see Fig.~\ref{fig:helix} (a). The vector rendering confirms the presence of a relaxed helical structure \citep{gray_prl} inside the SFC. As time progresses, we start seeing a second lobe as well. Two pairs of $\dot{B}$ probes are located outside the SFC to see if there is any magnetic flux leakage through the hole at the end wall. Substantially lower signal from these two probe pairs indicates that little flux leaks through the back end of the SFC. 

From the magnetic fluctuation spectrum of the relaxed Taylor state in the compression volume, we found that the spectrum is dominated by energy at low frequencies and long wavelengths. The spectral index is much steeper for the relaxed object ($\alpha < -4$) than that observed in turbulence studies during relaxation ($\alpha = -2.47$) \citep{Schaffner2014}. Our primary objective in the present work is to study the additional heating of the fully-relaxed Taylor state due to compression.
\subsection{Determination of local compression using wavelet analysis}
We determine the fractional compression of the relevant volume of plasma by measuring the changes in the dominant axial wavenumber of the magnetic structure.
Because the plasma is well-described by a helical analytic equilibrium, i.e., a Taylor state \citep{gray_prl}, it is possible to map changes in the local axial wavenumber of the helix directly to changes in the length of a parcel of plasma.
We use a continuous wavelet transform of the multi-point magnetic signals to calculate the spatial wavenumber spectrum as a function of 
time and space.
The wavelet transform is analogous to a Fourier transform, but uses basis functions with a finite extent in the transform variable \citep{Torrence_Compo_1998}.
We employed a Paul wavelet of order 2 as our mother wavelet function because it matches the qualitative spatial features of the magnetic field in the SFC.
Using a wavelet method (as opposed to windowed Fourier transforms) allows us to localize the power (in $z$) in an optimal way for each wavenumber, providing a compression measurement specific to the region where the temperature and density are monitored.

\begin{figure}
	\includegraphics[width=0.75\textwidth]{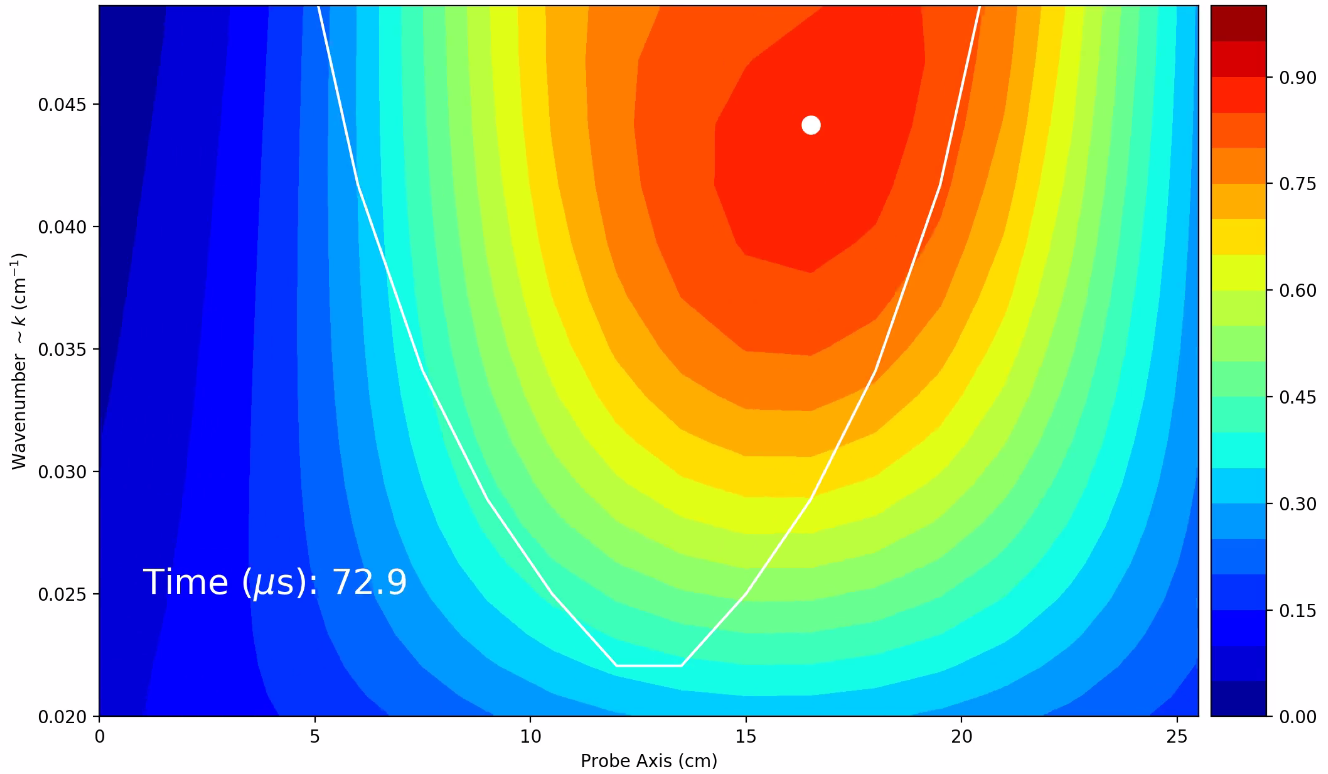}
	\caption{\label{fig:wavelet} The image of the wavelet analysis at a particular time instance for the same shot as Fig.~\ref{fig:plasma_param}. The colorbar on the right hand side tells the power in a particular wavenumber. The white dot represents the centroid of the most dominant wavenumber and the hyperbola-like white curve represents the cone of influence below which the wavenumber values are unreliable.}
\end{figure}

At each time step, we calculate a spatial wavelet spectrum of one component of the magnetic field transverse to the axis of the helix and one such wavelet spectrum is shown in Fig.~\ref{fig:wavelet}%
. 
For illustrative purposes, let $B^t(z) = B_x(t=t_i,z)$ for a particular time, $t_i$. 
There are 18 magnetic probe locations associated with the SFC, each spaced by $\Delta z = 1.5\,\textrm{cm}$ and indexed by the labels $m$ and $m'$.
Using this notation, $B^t(z) = B^t(m'\Delta z) = B^t_{m'}$ is the field component at position index $m'$ at time $t_i$.
The spatial wavelet transform at that time is:
\begin{equation}
W^t_m(s) = \sum_{m'=0}^{N-1} B^t_{m'}\Psi^*\left[\frac{(m - m')\Delta z}{s}\right]	,
\end{equation}  
where $\Psi$ is the mother wavelet function, $s$ is the wavelet scale (whose relationship to $k_z$ depends on the particular mother wavelet), $N = 18$ is the number of measurement locations, and $m$ represents the index of the spatial measurement point corresponding to the center of the wavelet. $W^t_m(s)$ thus represents the component of the spatial magnetic structure at time $t_i$ represented by a wavelet of scale $s$ centered at $m\Delta z$.

By tracking the peak of the power spectrum, $|W^t_m(s)|^2$, represented by white dot in Fig.~\ref{fig:wavelet}, we identify the dominant scale as a function of time.
As the scale changes, we can identify local changes in the pitch of the helical equilibrium structure corresponding to compression or rarefaction along the axis of the device.
The fractional change in volume is simply related to the fractional change in length by the assumption that the plasma fills the device radially: $\delta V/V =A\delta L/AL = \delta L/L$.
The typical wavenumber of the twist in the equilibrium state is $k_z\sim 0.04~cm^{-1}$, corresponding to a wavelength of $\lambda_z\sim 25~cm$, and typical changes in volume corresponding to compression are of the order of $10-30\%$. We found that the observed pitch of the Taylor state from axial $\dot{B}$ probe array is consistent with earlier more detailed measurements \citep{gray_prl}.

\subsection{PV diagram}
\begin{figure}
	\includegraphics[width=0.75\textwidth]{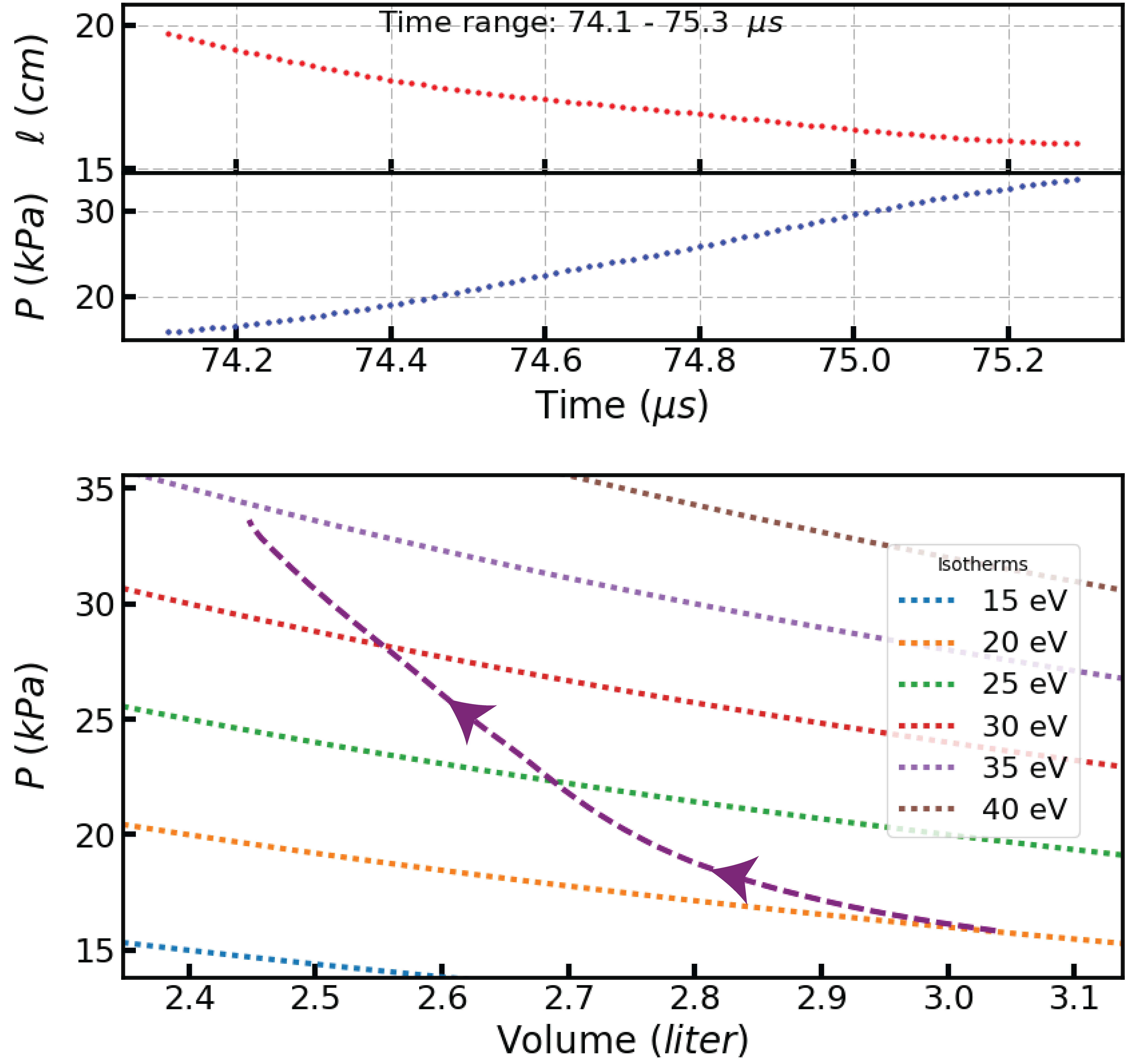}
	\caption{\label{fig:PV_dia} (a) A time trace of effective length (for the same shot as Fig.~\ref{fig:plasma_param}) of the Taylor state, (b) an associated increase in pressure, and (c) a trajectory of the thermodynamic state of the object in a PV diagram for the compression process. Uncertainty in the thermal pressure is $\approxeq15\%$ and is determined by propagating the errors in density and ion temperature. The uncertainty in the length of the Taylor state is $\approx10\%$ and is determined using the innermost contour of the wavelet analysis (see Fig.~\ref{fig:wavelet}). Note that as the volume of the Taylor state decreases the pressure shifts to higher isotherms.}
\end{figure}

The PV diagram describes the relationship between pressure and volume during any physical process. Using time traces of Taylor state length and plasma pressure (which is the product of plasma density and temperature), we construct a PV diagram. In the PV diagram, we identify the time windows which correspond to ion heating during compression. In other words, when compression and ion heating occur simultaneously, we call this a compressive heating event. A typical PV diagram for one such compression event is shown in Fig.~\ref{fig:PV_dia}.%
The PV diagram clearly shows that as the Taylor state compresses, the plasma pressure increases, resulting in a shift from a lower to higher isotherm in the PV diagram. In these experiments, a maximum Taylor state length compression of up to $30\%$ is observed. 

\subsection{Equations of state}
After the identification of a compressive heating event, the EOS of the compressed plasma are computed to assess the thermodynamics of the magnetized plasma. Because SSX plasmas relax to an equilibrium described by MHD \citep{cothran_prl, gray_prl}, one might expect the thermodynamics to be described by the corresponding EOS. To check the validity of equations of state, hundreds of shots are recorded under a variety of gun parameters. In all these shots, compressive heating events are identified on a PV diagram. The compression events, for most of the shots, range from $1 - 2~ \mu s$ in duration.

\begin{figure}
	\includegraphics[width=0.75\textwidth]{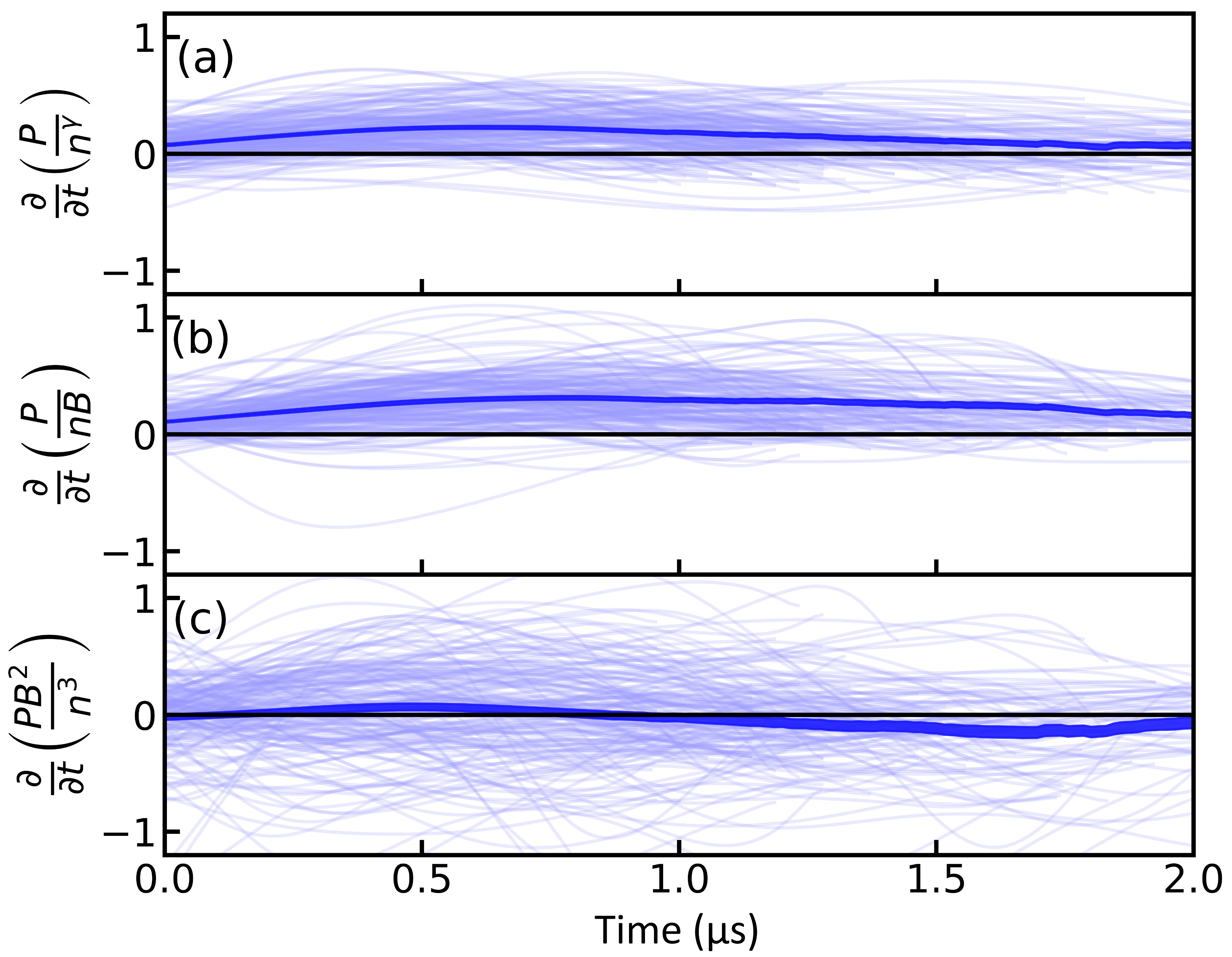}
	\caption{\label{fig:eos} Statistical variation of three equations of state for 192 compression events (length contraction ranges from $10 - 30\%$) are shown by light blue lines for: (a) the magnetohydrodynamic equation of state for 3D compression ($\gamma = 5/3$), (b) perpendicular, and (c) parallel CGL equations of state as a function of time. The dark blue band in each panel shows the standard error of the mean. Note that the MHD equation of state (a), and the perpendicular version of the CGL EOS (b) have nonzero time derivative. However, the parallel version of the CGL EOS (c) has nearly zero time derivative for most the compression time. Reprinted with permission from M. Kaur et. al., Phys. Rev. E 97, 011202 (2018). Copyright 2017 The American Physical Society.}
\end{figure}

To determine the general behavior of the compressed states, we plot the EOS for all the events in Fig.~\ref{fig:eos}.%
A set of 192 compression events are used for analyzing the EOS with a typical length contraction ranging from $10 - 40\%$ and are shown by light blue lines in each panel of Fig.~\ref{fig:eos}. The MHD EOS for 3D compression ($\gamma = 5/3$), perpendicular, and parallel CGL EOS as a function of time are plotted in Fig.~\ref{fig:eos} $(a)$, $(b)$ and $(c)$, respectively. The dark blue band in each panel shows the mean with the standard error of the mean represented by the thickness of the band.  
In Fig.~\ref{fig:eos}, note that the mean of the time derivatives specified by Eq. (\ref{mhd_eos}) - (\ref{<cgl_perp>}) are clearly nonzero, which indicate that our plasma does not satisfy the MHD and perpendicular CGL EOS. Whereas, the mean of the time derivative specified by Eq. (\ref{<cgl_par>}) remains nearly zero during compression which indicates that the parallel CGL EOS is valid for our magnetized plasmas. 

\begin{figure}
	\includegraphics[width=0.75\textwidth]{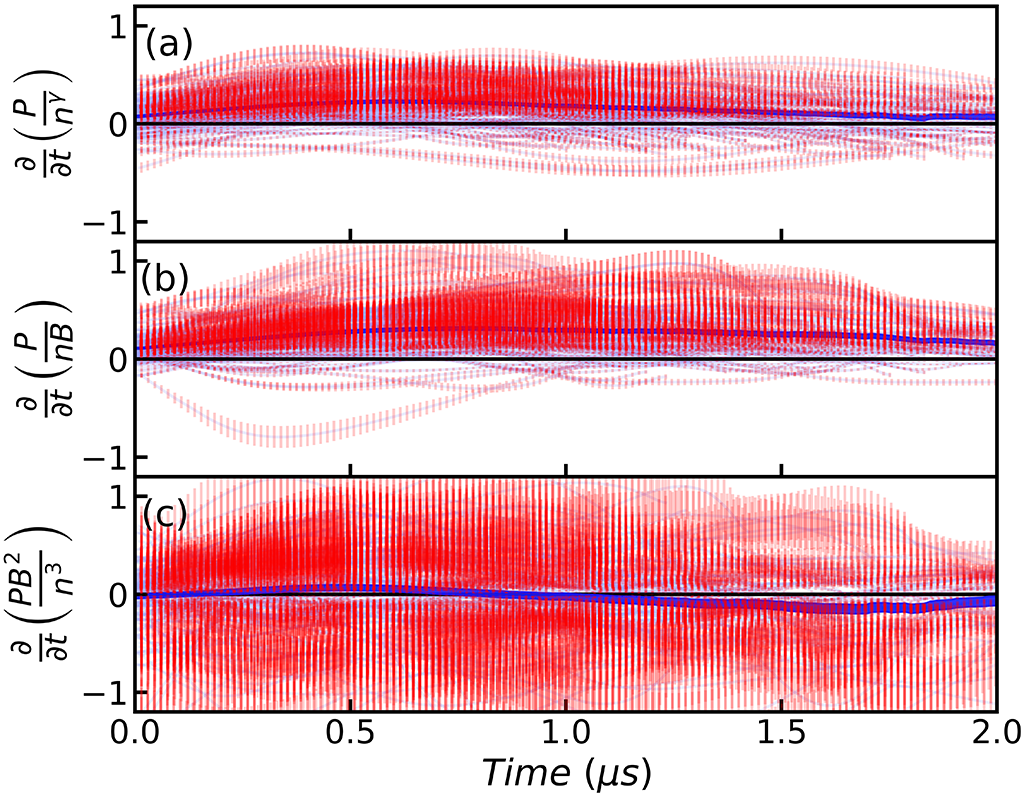}
	\caption{\label{fig:eos_error} Statistical time variation of three equations of state for 192 compression events is shown by the light blue lines for: (a) the MHD EOS for 3D compression, (b) perpendicular, and (c) parallel CGL EOS. The dark blue band in each panel shows the standard error of the mean, while the red color bars represent the errorbars in each individual event. It is clear that the errorbars for the MHD EOS (a), and the perpendicular version of the CGL EOS (b) are not consistent with the zero axis. However, the parallel version of the CGL EOS (c) errorbar is evenly distributed around zero axis.}
\end{figure}

The uncertainty in density and magnetic field is $\sim10\%$ and the error for $T_i $ is shown in second panel of Fig.~\ref{fig:plasma_param}. When we reconstruct Fig.~\ref{fig:eos} along with the individual event errorbars for all the three EOS, as shown in Fig.~\ref{fig:eos_error},%
the error-bars for MHD EOS and perpendicular CGL EOS are not consistent with the zero-axes. Whereas, the error-bars for parallel CGL EOS is evenly distributed around the zero-axis.   
We, further, note that our plasma is weakly-collisional so that could be the reason why the MHD EOS is not valid for our magnetized plasma. As the plasma ion-ion collisionality reduces ($\omega_{ci}\tau = 0.6$), the magnetic field starts to dominate. Because of this, the CGL EOS with proton energy in the parallel direction with respect to the magnetic field seems to describe the behavior of our plasmas.   

All the compression events used to construct Fig.~\ref{fig:eos} and Fig.~\ref{fig:eos_error} satisfy the following criteria: ($i$) Length compression should be greater than or equal to $10~\% $, ($ii$) compression event should last for more than $1~\mu s$, and ($iii$) the PV diagram should demonstrate a transition from a lower to higher isotherm. 
We treat these events as adiabatic since the time scale is of the order of one ion cyclotron time. 
Before taking the time derivative, quantities in each panel are normalized using their respective maxima such that the unit of each EOS in Fig.~\ref{fig:eos} is inverse of time. Since we measure the ion temperature along a chord, our $T_{i}$ measurements are insensitive to the direction of magnetic field. We are, therefore, obliged to use total pressure to test the various EOS instead of using $p_{\bot}$ and $p_{\parallel}$ separately. However, using the total pressure for the EOS analysis, we have noted that the parallel CGL EOS is most consistent with our data, which suggests that our plasma consists of mostly parallel proton energy. We would also like to mention that we are not trying to verify that one model or another is correct, but rather noting that one model fits our data best. 

\section{Simulation results:}
Because we use minimally invasive diagnostics for the measurement of plasma parameters in an azimuthal plane, we have no way of eliminating the possibility of profile modifications during compression. 
In order to get information about the profile modifications at other axial locations, we employ simulation techniques. Therefore, the main purpose of simulations is to explore the nature of axial and radial variation of the plasma parameters.  

\begin{figure}
	\includegraphics[width=\textwidth]{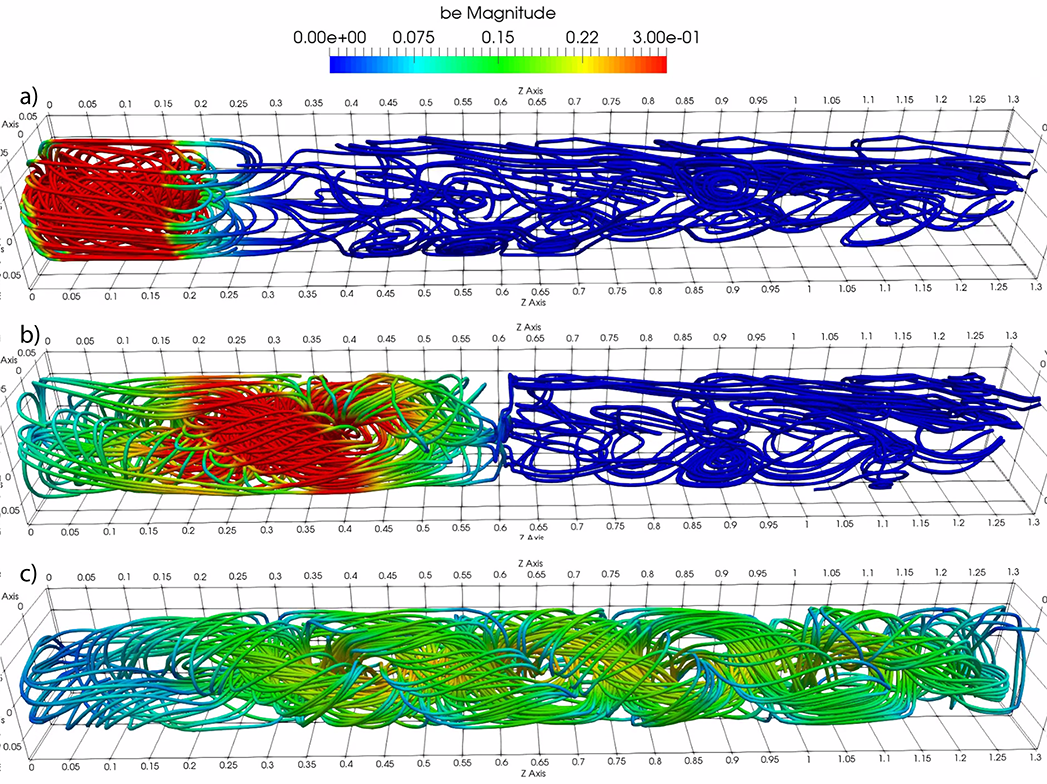}
	\caption{\label{fig:3D_rendering} The sequence of formation of a Taylor state: (a) In the beginning, a spheromak is formed at the left end, (b) at $2~\mu s$, it tilts and starts relaxing to a helical magnetic flux ropes structure, and (c) shows a relaxed, twisted Taylor state at $6.4~\mu s$. Due to its inertia, The relaxed structure moves towards the right hand side and fills almost the full volume.}
\end{figure}

We are employing a 3D resistive MHD code (NIMROD \citep{Sovinec2004}) to simulate the twisted Taylor states, which possess helical magnetic field lines resembling a rope (called a flux rope). One such helical structure obtained from these simulations is shown in Fig.~\ref{fig:3D_rendering}.
The simulations are carried out on a $32$ radial $\times$ $192$ axial grid with $4^{th}$ degree polynomials as the basis functions, running with three modes in the azimuthal direction. The boundary conditions are as follows: $1)$ The grid is specified as rectangular and non-periodic in axial direction, $z$, $2)$ The first wall is assumed to be a perfectly conducting boundary with $B_{\perp}=0$, $3)$ a `no-slip' boundary condition was assumed for velocity, i.e., $v=0$, and $4)$ Dirichlet conditions for density.
The initial conditions are as follows: $1)$ an axial velocity, $v_z = 40~km/s$ is applied to a CT that is initialized using a Bessel function model to resemble a spheromak with $2)$ uniform density of $n=1\times10^{20} ~m^{-3}$, and $3)$ edge poloidal magnetic field, $B=1~T$. In our simulations, Ohm's law does not include the Hall terms. The continuity advance is fully three-dimensional. Thermal conduction is isotropic and is not temperature dependent.

In Fig.~\ref{fig:3D_rendering}, a sequence of formation of a twisted Taylor state from a spheromak is shown. The color bar on the top of this figure represents the strength of the background magnetic field. 
After production, the spheromak relaxes to a twisted structure in a few Alfv\'en times. A high value of density and a low magnetic field would result in a long Alfv\'en time, $\tau_A \propto \sqrt{n}/B$. Therefore in order to reduce the cost of the simulations, we have used a higher value of magnetic field and a smaller value of density than is measured to shorten the simulation relaxation time. The results obtained in this way will be helpful in predicting structural behavior of our plasma.   

\begin{figure}
	\includegraphics[width=\textwidth]{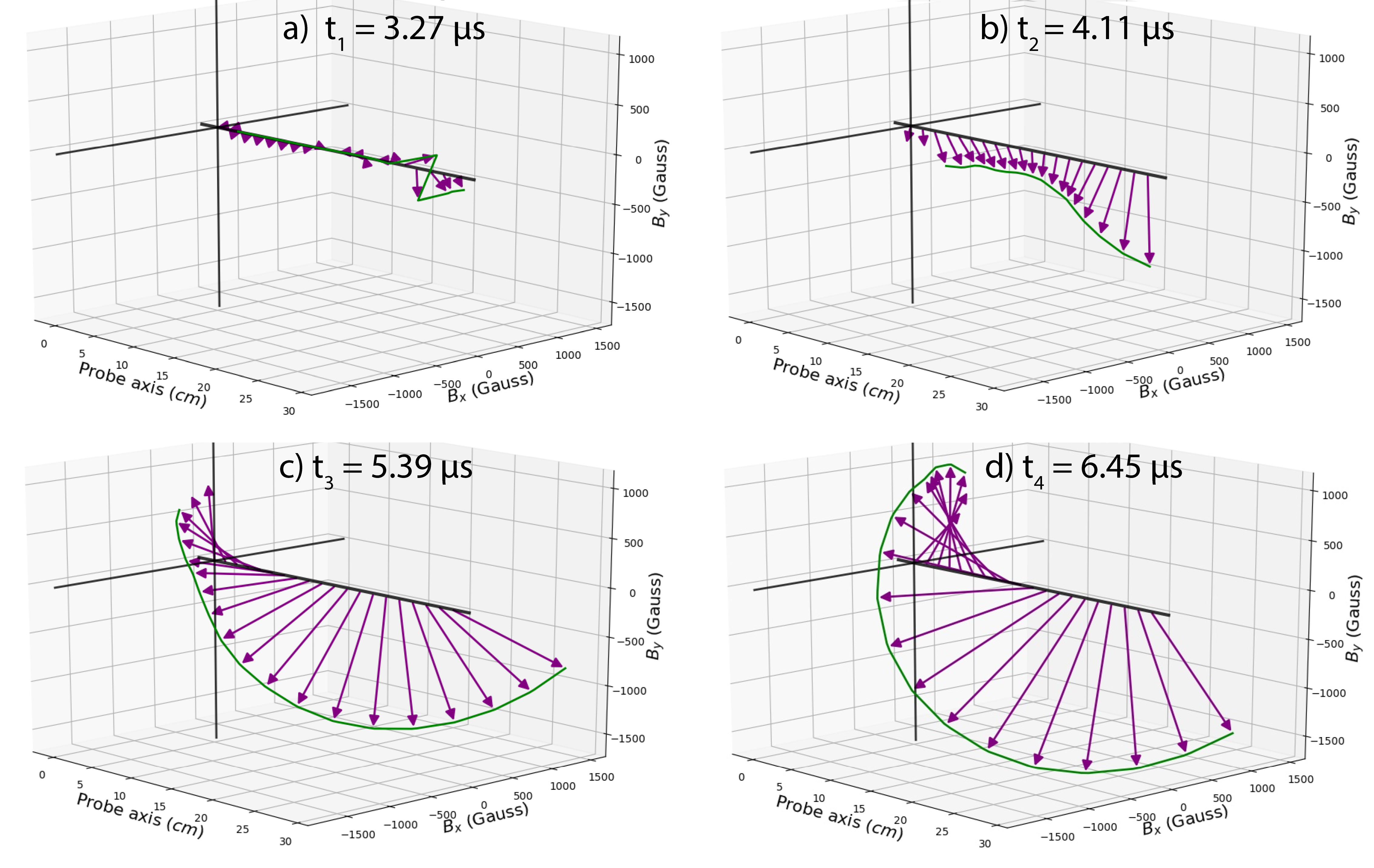}
	\caption{\label{fig:2D_rendering} Simulation rendering of magnetic field vectors along the axis of the SFC at four different times relative to the start of the simulation: (a) $t_1 = 3.27~\mu s$ corresponds to the time when Taylor state enters the mouth of the SFC, (b) $t_2 = 4.11~\mu s$ corresponds to when the Taylor state reaches almost midway, (c) $t_3 = 5.39~\mu s$ when Taylor state reaches the end wall, and (d) $t_4 = 6.45~\mu s$ corresponds to the time when Taylor state gets compressed against the end wall of the SFC. The two solid lines represent the end wall of the SFC and the Taylor state enters from the right hand side. These vector renderings are similar to what we observe in experiments and shown in Fig.~\ref{fig:helix}}
\end{figure}

Fig.~\ref{fig:2D_rendering}%
shows a magnetic field vector rendering along the axis of the SFC at four different times. The locations of synthetic magnetic diagnostics are chosen to replicate the actual magnetic probes in our experiments. The simulation results are similar to what we observe in our experiments, see Fig.~$\ref{fig:helix}$. As we mentioned earlier, there is a considerable difference in the magnitude of magnetic field in comparison to experiments but their overall behavior is similar to our experimental results. Initially in Fig.~\ref{fig:2D_rendering}, we see one helical lobe of the Taylor state inside the SFC and after some time, a second lobe also enters the SFC. 

\begin{figure}
	\includegraphics[width=0.75\textwidth]{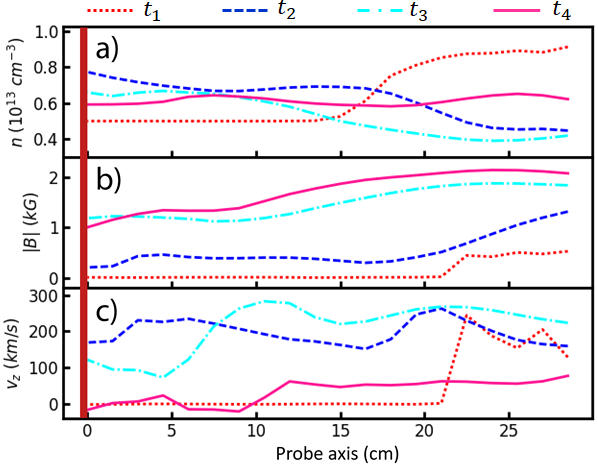}
	\caption{\label{fig:simu_parameters} Axial variation of the Taylor state (a) plasma density, (b) magnetic field, and (c) axial velocity from NIMROD simulations at different times. Vertical red bar in each panel represents the back end of the SFC, therefore  $29~cm$ corresponds to the beginning of the SFC. Different colors in each panel correspond to the axial values at three different time instances (same as in Fig.~\ref{fig:2D_rendering}); $i$) the dotted red line corresponds to time $t_1 = 3.27~\mu s$, $ii$) the dashed blue line to $t_2 = 4.11~\mu s$, $iii$) the dash-dot cyan line to $t_3 = 5.39~\mu s$, and $iv$) the pink solid line corresponds to $t_4 = 6.45~\mu s$. It can be clearly seen that the Taylor state velocity decreases to zero during compression near the back end of the SFC.}
\end{figure}

In Fig.~\ref{fig:simu_parameters},%
plasma density, the magnetic field embedded in the Taylor state, and its axial velocity are plotted at the same four time instances corresponding to the vector rendering of magnetic field in the SFC in Fig.~\ref{fig:2D_rendering}. The Taylor state enters the SFC from right hand side and the red vertical bar indicates the end of the SFC. The simulation results demonstrate an amplification in the magnetic field as the Taylor state compresses against the end wall of the SFC. However, the plasma density is not observed to change much near the end wall during compression. The drop in axial velocity indicates that the Taylor state stagnates after reaching the end wall of the SFC. 

\section{Discussion:}
We have identified compressional heating events by constructing PV diagrams in an MHD plasma. During these events, we observe a maximum length contraction of a factor of $30 - 40\%$. The time windows of 192 compression events are used for analyzing various equations of state applicable for magnetized plasmas. 

The equilibrium of these plasmas is described by MHD but the EOS analysis shows that the MHD EOS does not describe the dynamics of our plasma. It may be because of the fact that our plasma is weakly-collisional; $ \omega_{ci}\tau=0.6$ ($\approxeq 1$) for average plasma density of $\sim2\times10^{15}~cm^{-3}$, ion temperature of $\sim15 ~eV$, and magnetic field of $\sim1500~G$. Due to the weak ion-ion interactions, the effect of magnetic field becomes more dominant on their trajectories.  

The perpendicular CGL EOS also does not hold. But surprisingly, the parallel CGL EOS better characterizes the average behavior of our plasma. Our line-averaged IDS measurements of ion temperature are insensitive to the direction of the local background magnetic field. Nonetheless, we see a difference between the perpendicular and parallel CGL EOS.

We hypothesize about the physical process in the following way. We produce a spheromak using the coaxial magnetized plasma gun. As the spheromak moves away from the gun, it expands and relaxes immediately after formation and the structure unravels towards a twisted Taylor state. In the process of relaxation, the magnetic field associated with the spheromak drops by a factor of 10 or more. Due to the conservation of the first adiabatic invariant, the reduction in the magnetic field causes a concomitant drop in $T_\bot$. By the time the Taylor state enters the stagnation flux conserver located at the other end of the linear device, we suspect that most of the ion energy is in $T_\parallel$. The compression events in the stagnation flux conserver reduce the volume by only 10 - 30\%, which is not enough to lead to a considerable increase in $T_\bot$. The expansion of the solar wind may involve similar physics \citep{Brown2015}. 

Impurities in the hydrogen plasma, through ionization and excitation states, may affect the $\gamma = 5/3$ assumption to MHD EOS. However, 
our past studies have shown that the impurity level in SSX is minimal 
\citep{Chaplin2009} and our plasma is optically thin so we do not expect the radiation to contribute to the EOS significantly. In any case, incorporating these effects is beyond the scope of this paper.

In summary, we present a PV analysis of the compression process of a magnetized, relaxed plasma leading to ion heating. A comparison of equations of state applicable to the compression process is carried out which suggests that the parallel  CGL equation of state models the average behavior of our plasma better than other equations of state. Although the magnetized plasma is well-described by an MHD equilibrium, the MHD EOS does not agree with experimental observations. Simulation results confirm the presence of a relaxed helical structure and also demonstrate Taylor state stagnation.  

While passing through the glass tube, the magnetized plasma can be accelerated using a pulsed magnetic field. The pulsed magnetic field is produced by discharging a capacitor through a pinch coil having a fast quarter cycle rise time. In this way, we wish to accelerate these Taylor states to higher velocity in our future experiments and then compress them to higher density in the SFC. These experiments have already started with one pinch coil. In the near future, we are planning to use four such coils. These coils will be triggered separately and sequentially to accelerate plasma to velocities over $200~km/s$, ion temperature over $100 ~eV$ and to achieve compressional density over $10^{16}~ cm^{-3}$ in the SFC. Expected higher compression in these acceleration experiments would help in the better understanding of the EOS, as would improved diagnostic coverage. 

\section*{Acknowledgment}
This work is supported by the Accelerating Low-
Cost Plasma Heating and Assembly (ALPHA) Program of
the Advanced Research Projects Agency–Energy (ARPA-E) and Office of Fusion Energy Sciences. We wish to acknowledge Steven Palmer and Paul Jacobs for their technical support. 

\bibliographystyle{jpp}

\bibliography{Dr_Kaur_JPP}

\end{document}